***Note**: Early results that matured into Chap 11 & 12 of Mathematical Geoenergy (Wiley/AGU 2018). Mathieu equation replaced with Laplace's Tidal Equations to represent the sloshing.*

# Sloshing Model for ENSO

Revised 8/28/2020 13:21:00

P.R. Pukite

*e-mail address: puk@umn.edu*

The El Nino southern oscillation (ENSO) behavior can be effectively modeled as a response to a 2nd-order Mathieu/Hill differential equation with periodic coefficients describing sloshing of a volume of water. The forcing of the equation derives from tidal-forcings as in QBO, angular momentum changes synchronized with the Chandler wobble, and a metastable biennial modulation. One regime change was identified in 1981.

## I. INTRODUCTION

A general sloshing formulation is modeled as the following 2nd-order differential wave equation [3][5][6]

$$f''(t) + \left(\omega^2 + q(t)\right) f(t) = F(t) \qquad (1)$$

where $f(t)$ is the level height and $F(t)$ is a forcing. For ENSO, the characteristic frequency is given by $\omega$, which has been evaluated as $2\pi/4.25$ rads/yr, based on dynamic thermocline behavior [1]. The factor $q(t)$ is a non-linear Mathieu or Hill-type modulation that arises as a natural consequence of a constrained volume of largely inviscid liquid [3], and can be further induced by a vertical forcing [6]. Although the physics of the sloshing behavior is ultimately complex, the more elaborate finite-element simulations remain close to the result of equation (1) if $q(t)$ and $F(t)$ are periodic functions [6]

The results of this study reveal that if $F(t)$ corresponds to a mixed forcing of the QBO long-period tidal factors, Chandler wobble, and a biennial modulation s, combined with a characteristic period of 4.25 years, a surprisingly good fit to the Southern Oscillation Index (SOI) time-series of ENSO is obtained. The SOI time series was chosen because it is well characterized [30] and functions close to the oscillating standing-wave dipole [12] that is characteristic of a sloshing behavior. It also has a long-running record dating back 130+ years collected from the Tahiti (+ pole) and Darwin (- pole) sites.

Although the SOI is a measure of atmospheric pressure, via the reverse barometric effect one can tie in ocean-level variations as a result of spatio-temporal sloshing to changes in pressure. This becomes the SOI Model (SOIM).

## II. QUASI-BIENNIAL OSCILLATION

The quasi-biennial oscillation (QBO) of stratospheric winds has long been associated with ENSO [7][8][14][27], and has been thought to produce a forced stimulation to the ocean's surface by a down-welling wind shear. The QBO cycles by exhibiting two longitudinal direction reversals every 28 months on average. Measurements for QBO at different altitudes (expressed as an equivalent atmospheric pressure) are available since 1953 [36]. The goal with fitting QBO was to find potential common forcing mechanisms.

Lindzen [23] first suggested that *"Lunar tides are especially well suited to such studies since it is unlikely that lunar periods could be produced by anything other than the lunar tidal potential."* Initially, this study applied a machine learning tool to isolate the primary QBO frequencies. **Table 1** shows that it discovered two frequencies corresponding to *seasonally aliased* lunar month periods (vs non-aliased [14]).

*Table 1: Draconic (1st row) and Tropical (2nd row) aliasing*

| aliased frequency | period | unaliased days | closest lunar | % error |
|---|---|---|---|---|
| **2.663410** | 2.359075 | 27.20835 | 27.21222 | -0.01424 |
| **2.297534** | 2.734752 | 27.32689 | 27.32158 | 0.019416 |

The wave model targeted the 30 hPa altitude measure of QBO as this showed the strongest signal-to-noise ratio. To confirm, we fit QBO using the seasonally aliased values of the draconic, anomalistic, and tropical long-period tides and extracted the 2nd-derivative to isolate faster aliased periods.

$$f''(t) \sim F(t) \qquad (2)$$

**FIG. 1** shows a multiple regression fit with a training interval from 1953 to 1986 and a validation interval from 1986 to 2015. Yellow indicates the rare poorly fit regions.

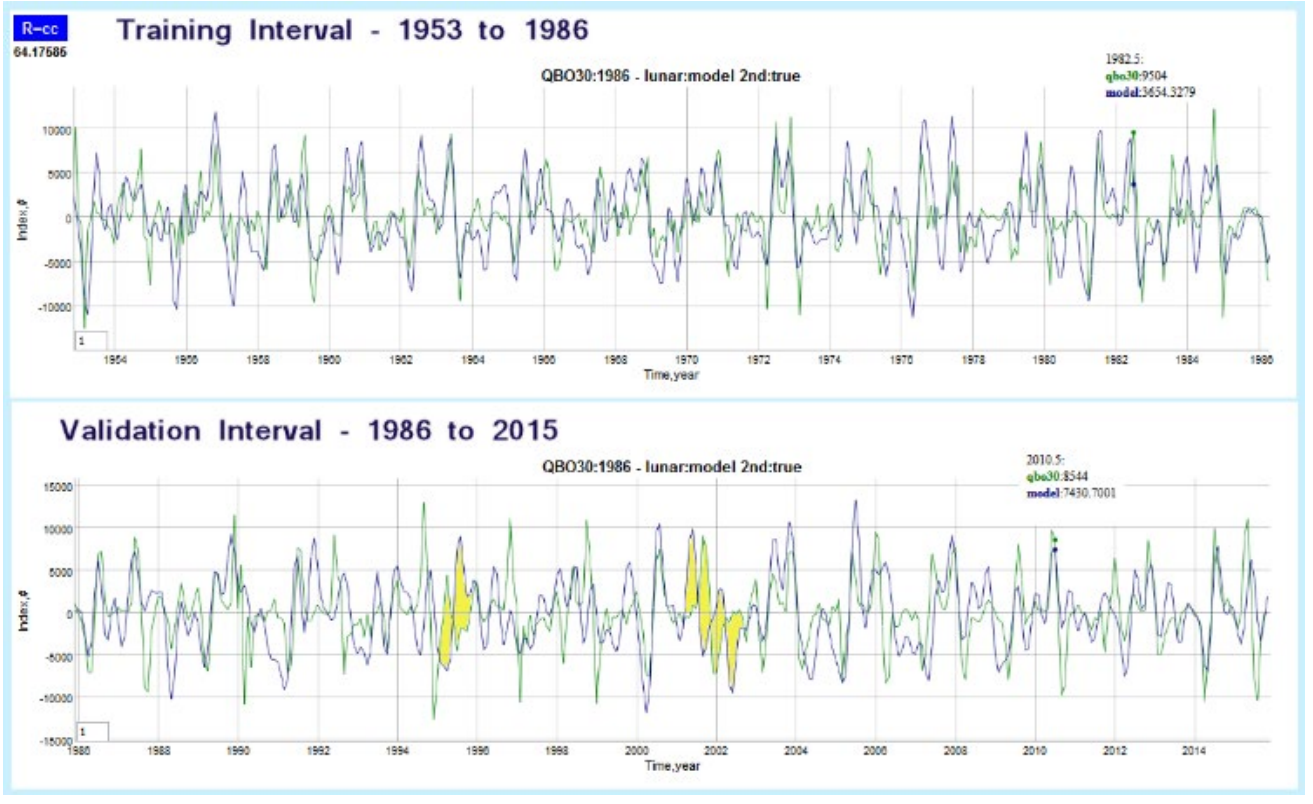

**FIG. 1**: *QBO model for wind velocity from year 1953*

The sensitivity of the fitting factors to the lunar long periods is shown in **Fig. 2** amidst the noisy 2nd-derivative.

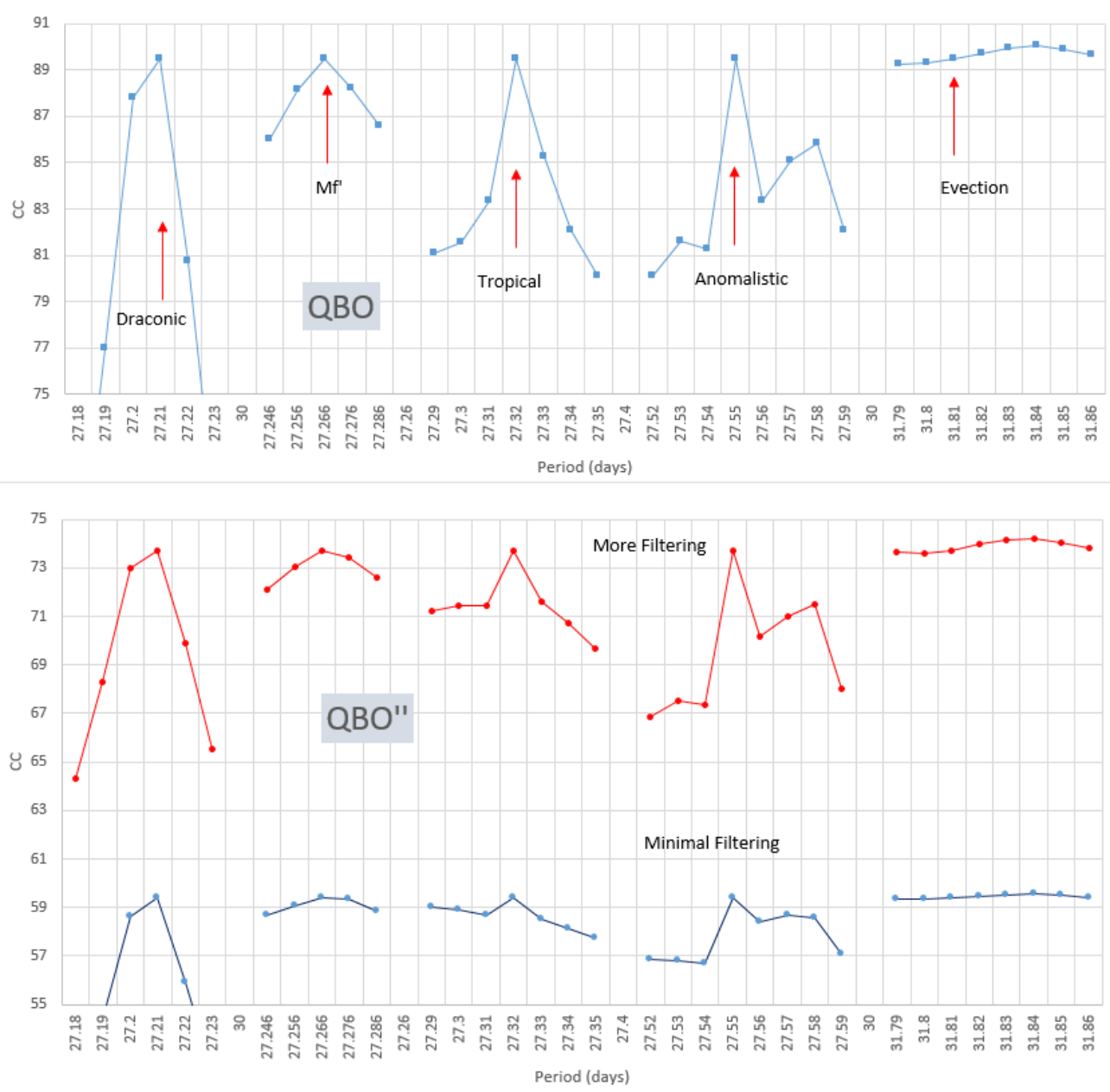

**FIG. 2**: *Correlation sensitivity of lunar tidal periods*

The relative strength of factors is shown in **Fig. 3**. In keeping with a wave equation model, the 2nd-derivative factors scale by $\omega^2$. In addition to the 30 hPa QBO series, the other altitude variants gave equally good fits, with the higher altitudes showing a greater semi-annual content.

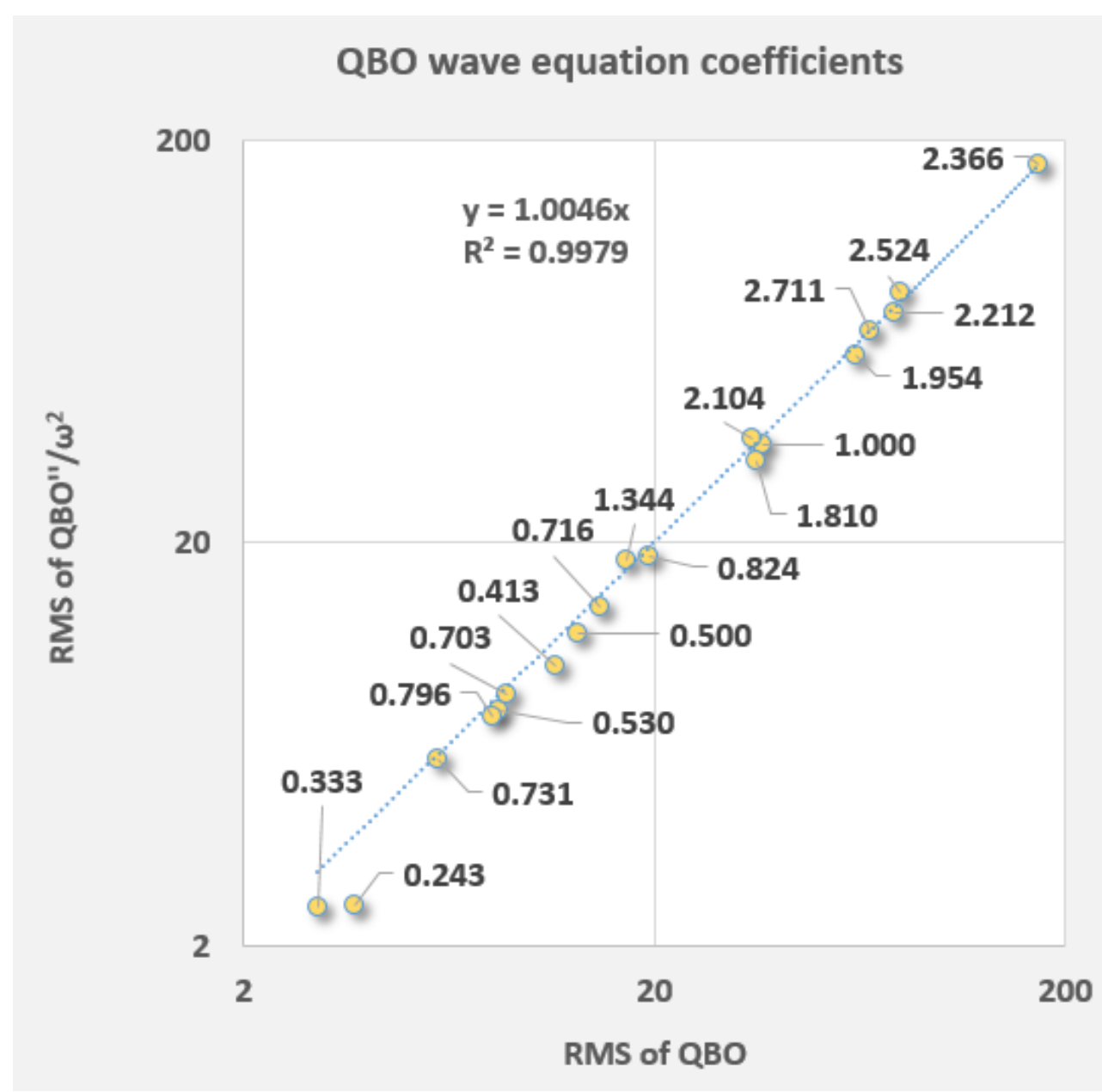

**FIG. 3**: *The strength of the aliased tidal period for modeled QBO and 2nd-derivative of QBO*

## III. CHANDLER WOBBLE (CW)

The geophysical mechanisms associated with the Chandler wobble in the Earth's rotation were first identified by Gross [9]. He proposed that fluctuating pressure in the ocean, caused by temperature and salinity changes and wind-driven perturbations in the ocean's circulation was a principle cause of the wobble. As this is considered a component of the conserved angular momentum of the earth's lithosphere, a Chandler wobble factor is included in the SOI model along with the same lunar-forcings found in the QBO. Because of the larger moment of inertia in the ocean, it is reasoned that additional angular momentum changes than that produced directly by cyclic lunar forcing would apply to ENSO. Of course, this would explain the greater cyclic variation in the ENSO time-series profile.

The measure of the Chandler wobble that would apply in this case is derived from measurements of the polar $x$ and $y$ coordinate velocity [31]

$$\dot{r} = \sqrt{\dot{x}^2 + \dot{y}^2}$$

The JPL POLE99 Kalman Earth Orientation Series filtered data set was used to model this quasi-periodic oscillation [34]. An average value of 6.46 years for the velocity period was estimated [21] while some findings suggest that the Chandler wobble is a split between closely separated spectral peaks [10]. The period of 6.46 years correlates with the beat frequency of the Chandler wobble period of between 432 and 433 days and the annual cycle (see **FIG. 4**).

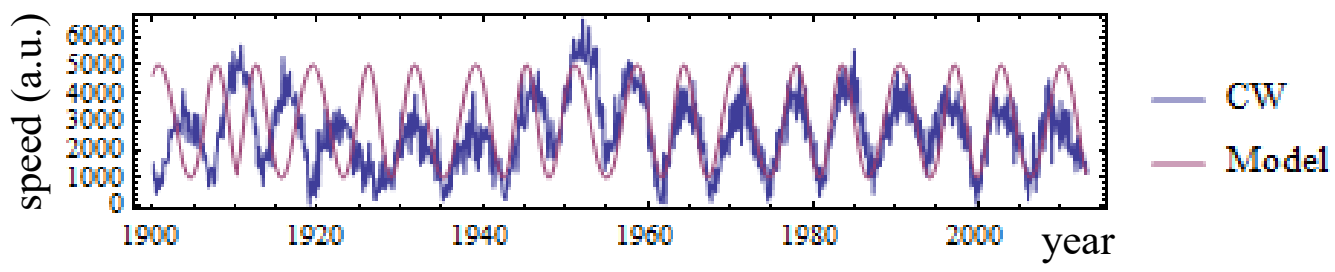

**FIG. 4**: *Chandler wobble model (y axis = arb. units for* $\dot{r}$*)*

## IV. TOTAL SOLAR IRRADIANCE (TSI)

Solar variations have been associated with ENSO [29] via its role in modulating the heating/cooling of the water volume. ENSO is considered a cyclic recharge/discharge pattern, so one could intuit that excess solar flux can reinforce a resonance condition if modulated at an appropriate rate. Direct correlations between the Schwabe/Hale cycles of solar flux variations and the ENSO pattern have yet to be found [11], but it may impact the QBO [18]. Although we can't rule out TSI modulation as a factor, introducing a ~11-yr period does not improve the fit.

## V. MATHIEU MODULATION

A biennial modulation is found to apply to the *q(t)* term in equation (1) of the sloshing wave equation. According to Frandsen [6], any vertical forcing to the RHS of the sloshing differential equation gives rise to a Mathieu-type modulation of the same frequency. We have found that a biennial or 2-year modulation agrees with that found by Dunkerton [4] and Remsberg [26] in stratospheric measurements and by Pan [24] in sensitive GPS of the earth's deformation. A strict biennial mode is also observed in ENSO measurements [13].

## VI. CLIMATE REGIME SHIFT

Any analysis of ENSO must consider the significant phase shift that started in 1976/1977 and lasted until the 1980's [20][30]. Marcus et al [17] analyzed angular momentum changes in the Earth's rotation and found a significant perturbation that they associated with the 1976/1977 shift. Astudillo *et al* [1] applied Takens embedding theorem (which works for linear and non-linear systems such as the Mathieu and Hill formulation) to the ENSO time series, reconstructing current and future behavior from past behavior and also found a strong discontinuity around 1981.

In fits to the sloshing model, the perturbation started at 1981 and lasted for 16 years. In terms of a biennial mode and metastable conditions there is no distinction between even or odd-year parity, so by inverting the phase of the biennial signal during that interval, a good fit was achieved.

## VII. COMPOSITION RESULTS

The factors described above were composited according to equation (1), and finalized as equation (2) below and evaluated via a differential equation solver, using SOI data collected since 1880 [35] (both *Mathematica* and *R* DiffEq solvers were used, with similar outcomes). The original forcing factor, *F(t)*, on the right-hand side (RHS) was formulated as a biennial factor multiplied by a mixed factor, *g(t)*, containing the lunar-tidal and Chandler wobble forcing.

$$f'' + (\omega^2 + a\cos(\pi t + \varphi))f = \cos(\pi t)\, g \quad (2)$$

Initially, a differential evolution search was attempted to optimize coefficients and phase terms, but a straightforward manual adjustment proved quicker. The *f(t)* term was compared to a combination of SOI and a fraction of the NINO3.4 time series to reduce noise. The correlation coefficient reached 0.8, which is likely close to a ceiling due to the noise differential among the Tahiti and Darwin [30]. The scaling was adjusted by equalizing the model and data variance. More noise was evident in the early years, where the model tracked Darwin data better than the spotty Tahiti.

An alternative wave-equation transformation approach to complement a direct differential equation solver was applied to the results. This approach applied the second-derivative to the LHS (left-hand-side) of equation (2) and compares directly to the RHS forcing. So what we see in **FIG. 6** is a biennial modulated forcing view of ENSO. In **FIG. 5** the excursions are compared after 1940. The term *g(t)* contains the Chandler wobble period of 6.48 years, a 14-year QBO-related term [26] and possibly triaxial wobble contribution [33], an 18.6 year draconic term, and a weaker anomalistic 4.065 year term, split into two aliased sidebands by the biennial modulation. The key to the excellent fit is applying a biennial phase inversion between the years 1981-1996, which effectively flips the even/odd year parity temporarily.

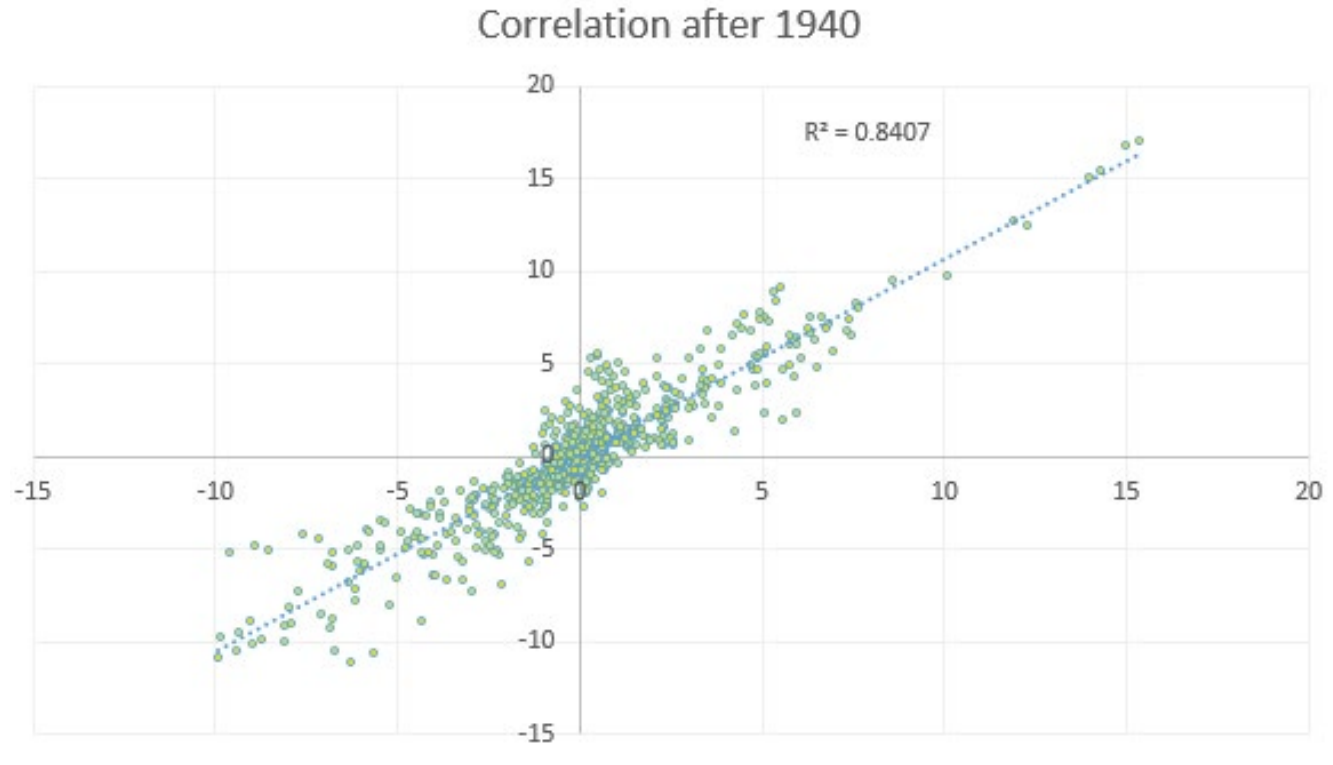

**FIG. 5**: *Scatter plot of ENSO excursions model vs data*

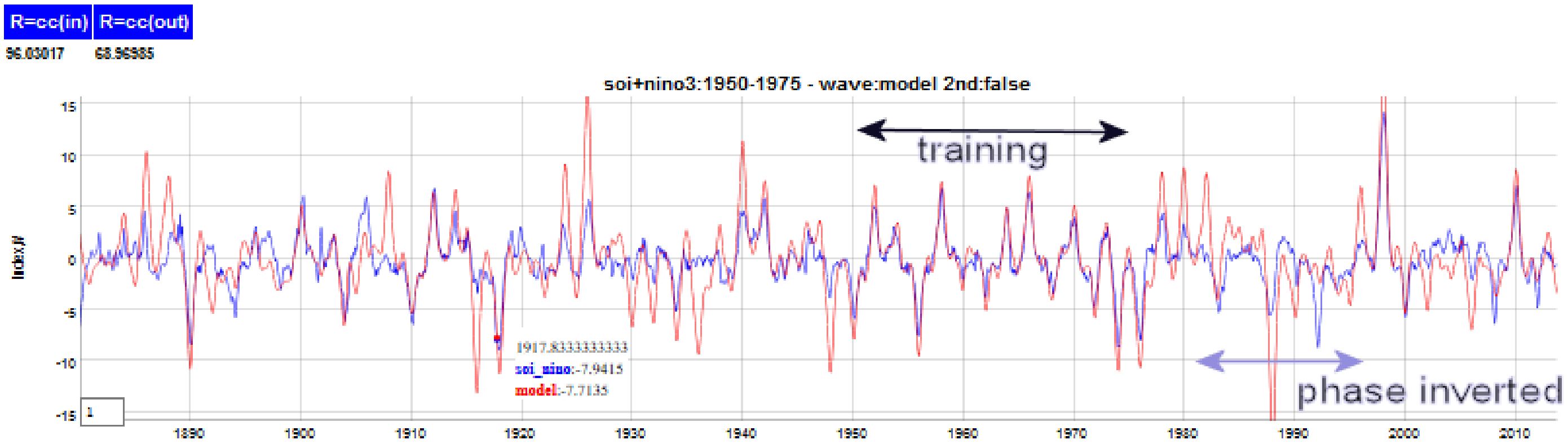

**FIG. 6**: *Aggregate model (SOIM) fit to the ENSO component via a wave equation transformation approach. The training interval is restricted to the years between 1950 and 1975..*

## VIII. DISCUSSION AND CONCLUSIONS

As a supplementary evaluation, the *Eureqa* symbolic regression machine learning tool was used to substantiate the selection of the forcing factors. The LHS of equation (3) was input to the *Eureqa* solver, where *D(soi,t,2)* is the second derivative of the SOI time series.

$$D(soi,t,2) + \omega^2 soi = F(t) \qquad (3)$$

Not surprisingly, the solver identified factors such as QBO and Chandler wobble in the *F(t)* expansion.

The model of ENSO as a forced response to a wave equation works remarkably well to reproduce historical data. It is well known that a periodic forcing can reduce the erratic fluctuations and uncertainty of a near-chaotic response function [22][32]. What the known tendency to locking of El Nino events towards the end of a calendar year [25] does is promote a biennial modulation *"its phase is strongly locked to the calendar months. Thus, the assumption of the 2-year periodicity seems reasonable"* [13]. The relative impact of El Nino events tend to be associated with the biennial periodicity. This model also treats the southern oscillation in its entirety, with the multiyear period forcing factors important in reproducing the detailed behavioral profile of the SOI time series.

Apart from the difficulty of predicting significant regime changes that can violate the stationarity requirement of the model, both hindcasting for evaluating paleoclimate ENSO data [19] and forecasting may have potential applications for this model. The model is simple enough in its formulation that others can readily improve in its fidelity without resorting to the complexity of a full blown global circulation model (GCM), and in contrast to those depending on erratic [28] or stochastic inputs which have less predictive power. Fits to a red noise model over a short training interval as shown in **Fig.6** will quickly diverge over the entire range since the coherence of the long term periodic forcing factors do not apply. So ENSO has a real potential for producing deterministic predictions years in advance.

---

[1] Astudillo, H., R. Abarca-del-Rio, and F. Borotto, "Long-term non-linear predictability of ENSO events over the 20th century," arXiv preprint arXiv:1506.04066, 2015.

[2] Clarke, Allan J, Stephen Van Gorder, and Giuseppe Colantuono. "Wind Stress Curl and ENSO Discharge/recharge in the Equatorial Pacific." *Journal of Physical Oceanography* 37, no. 4 (2007): 1077–91.

[3] Dubois, François, and Dimitri Stoliaroff. "Coupling Linear Sloshing with Six Degrees of Freedom Rigid Body Dynamics." *arXiv Preprint arXiv:1407.1829*, 2014.

[4] Dunkerton, Timothy J. "Quasi-biennial and subbiennial variations of stratospheric trace constituents derived from HALOE observations." Journal of the Atmospheric Sciences58.1 (2001): 7-25.

[5] Faltinsen, Odd Magnus, and Alexander N Timokha. *Sloshing*. Cambridge University Press, 2009.

[6] Frandsen, Jannette B. "Sloshing Motions in Excited Tanks." *Journal of Computational Physics* 196, no. 1 (2004): 53–87.

[7] Gray, William M, John D Sheaffer, and John A Knaff. "Hypothesized Mechanism for Stratospheric QBO Influence on ENSO Variability." *Geophysical Research Letters* 19, no. 2 (1992): 107–10.

[8] ———. "Influence of the Stratospheric QBO on ENSO Variability." *J. Meteor: Soc. Japan* 70 (1992): 975–95.

[9] Gross, Richard S. "The Excitation of the Chandler Wobble." *Geophysical Research Letters* 27, no. 15 (2000): 2329–32.

[10] Guo, JY, H Greiner-Mai, L Ballani, H Jochmann, and CK Shum. "On the Double-Peak Spectrum of the Chandler Wobble." *Journal of Geodesy* 78, no. 11–12 (2005): 654–59.

[11] Haam, Eddie, and Ka-Kit Tung. "Statistics of Solar cycle–La Nina Connection: Correlation of Two Autocorrelated Time Series." *Journal of the Atmospheric Sciences* 69, no. 10 (2012): 2934–39.

[12] Kawale, Jaya, Stefan Liess, Arjun Kumar, Michael Steinbach, Auroop R Ganguly, Nagiza F Samatova, Fredrick HM Semazzi, Peter K Snyder, and Vipin Kumar. "Data Guided Discovery of Dynamic Climate Dipoles," 30–44, 2011.

[13] Kim, Jinju, and Kwang-Yul Kim. "The tropospheric biennial oscillation defined by a biennial mode of sea surface temperature and its impact on the atmospheric circulation and precipitation in the tropical eastern Indo-western Pacific region." Climate Dynamics, 2016: 1-15.

[14] Li, G., Zong, H., & Zhang, Q. (2011). 27.3-day and average 13.6-day periodic oscillations in the Earth's rotation rate and atmospheric pressure fields due to celestial gravitation forcing. Advances in Atmospheric Sciences, 28, 45-58.

[15] Liess, Stefan, and Marvin A Geller. "On the Relationship between QBO and Distribution of Tropical Deep Convection." *Journal of Geophysical Research: Atmospheres (1984–2012)* 117, no. D3 (2012).

[16] Lindzen, Richard S., and Siu-Shung Hong. "Effects of mean winds and horizontal temperature gradients on solar and lunar semidiurnal tides in the atmosphere." Journal of the Atmospheric Sciences 31.5 (1974): 1421-1446.

[17] Marcus, Steven L, Olivier de Viron, and Jean O Dickey. "Abrupt Atmospheric Torque Changes and Their Role in the 1976–1977 Climate Regime Shift." *Journal of Geophysical Research: Atmospheres (1984–2012)* 116, no. D3 (2011).

[18] Mayr, Hans G, John G Mengel, and Charles L Wolff. "Wave-driven Equatorial Annual Oscillation Induced and Modulated by the Solar Cycle." *Geophysical Research Letters* 32, no. 20 (2005).

[19] McGregor, S., A. Timmermann, and O. Timm. "A Unified Proxy for ENSO and PDO Variability since 1650." *Clim. Past* 6, no. 1 (January 5, 2010): 1–17. doi:10.5194/cp-6-1-2010; Cobb, Kim M, Christopher D Charles, Hai Cheng, and R Lawrence Edwards. "El Nino/Southern Oscillation and Tropical Pacific Climate during the Last Millennium." *Nature* 424, no. 6946 (2003): 271–76.

[20] Miller, Arthur J, Daniel R Cayan, Tim P Barnett, Nicholas E Graham, and Josef M Oberhuber. "The 1976–77 Climate Shift of the Pacific Ocean." *Oceanography* 7, no. 1 (1994): 21–26.

[21] Miller, N, and Z Malkin. "Analysis of Polar Motion Variations from 170-Year Observation Series." *arXiv Preprint arXiv:1304.3985*, 2013.

[22] Osipov, Grigory V, Jürgen Kurths, and Changsong Zhou. *Synchronization in Oscillatory Networks*. Springer, 2007.

[23] Lindzen, Richard S., and Siu-Shung Hong. "Effects of mean winds and horizontal temperature gradients on solar and lunar semidiurnal tides in the atmosphere." Journal of the Atmospheric Sciences 31.5 (1974): 1421-1446.

[24] Pan, Yuanjin, et al. "The Quasi-Biennial Vertical Oscillations at Global GPS Stations: Identification by Ensemble Empirical Mode Decomposition." Sensors 15.10 (2015): 26096-26114.

[25] Rasmusson, Eugene M, Xueliang Wang, and Chester F Ropelewski. "The Biennial Component of ENSO Variability." *Journal of Marine Systems* 1, no. 1 (1990): 71–96.

[26] Remsberg, Ellis E. "Methane as a diagnostic tracer of changes in the Brewer–Dobson circulation of the stratosphere." Atmospheric Chemistry and Physics 15.7 (2015): 3739-3754.

[27] Taguchi, M. "Observed Connection of the Stratospheric Quasi-biennial Oscillation with El Niño–Southern Oscillation in Radiosonde Data." *Journal of Geophysical Research: Atmospheres (1984–2012)* 115, no. D18 (2010).

[28] Thual, Sulian, Boris Dewitte, Nadia Ayoub, and Olivier Thual. "An Asymptotic Expansion for the Recharge-Discharge Model of ENSO." *Journal of Physical Oceanography* 43, no. 7 (2013): 1407–16.

[29] Tiwari, RK, Rekapalli Rajesh, and B Padmavathi. "Evidence for Nonlinear Coupling of Solar and ENSO Signals in Indian Temperatures During the Past Century." *Pure and Applied Geophysics*, 2014, 1–13.

[30] Trenberth, Kevin E, Julie M Caron, David P Stepaniak, and Steve Worley. "Evolution of El Niño–Southern Oscillation and Global Atmospheric Surface Temperatures." *Journal of Geophysical Research: Atmospheres (1984–2012)* 107, no. D8 (2002): AAC – 5.

[31] Vondrak, Jan, Cyril Ron, and Vojtěch Štefka. "Earth Orientation Parameters Based on EOC-4 Astrometric Catalog." *Acta Geodyn. Geomater, in Print*, 2010.

[32] Wang, Geli, Peicai Yang, and Xiuji Zhou. "Nonstationary Time Series Prediction by Incorporating External Forces." *Advances in Atmospheric Sciences* 30 (2013): 1601–7.

[33] Wang, Wen-Jun, W.-B. Shen, and H.-W. Zhang, "Verifications for Multiple Solutions of Triaxial Earth Rotation," IERS Workshop on Conventions Bureau International des Poids et Mesures (BIPM), Sep. 2007.

[34] ftp://euler.jpl.nasa.gov/keof/combinations/2012/pole2012.pm

[35] http://www.cgd.ucar.edu/cas/catalog/climind/soi.html

[36] http://www.geo.fu-berlin.de/met/ag/strat/produkte/qbo/qbo.dat

[37] http://lasp.colorado.edu/data/sorce/tsi_data/TSI_TIM_Reconstruction.txt